\documentclass{article}
\usepackage{spconf,amsmath,graphicx}
\usepackage{cite}
\usepackage{amsmath,amssymb,amsfonts}
\usepackage{algorithmic}
\usepackage{graphicx}
\usepackage{textcomp}
\usepackage{xcolor}
\usepackage{booktabs}
\usepackage{subcaption}
\usepackage{dsfont}

\title{Unsupervised Data-Efficient Cross-Modal Retrieval with Global-Neighborhood Alignment Hashing}
\name{
    \begin{tabular}{c}
    Runhao Li$^{1}$, Xiaoxu Ma$^{2}$, Zhenyu Weng$^{2,*}$, Yue Zhang$^{3}$, Guibo Luo$^{4}$,\\
    Huiping Zhuang$^{2}$, Zhiping Lin$^{1}$, Yap-Peng Tan$^{5,1}$
    \thanks{
    This work was supported in part by Guangdong Provincial Key Laboratory of Ultra High Definition Immersive Media Technology (2024B 1212010006), Guangdong Basic and Applied Basic Research Foundation (Grant No. 2026A1515030047), Guangdong Grant No. 2024QN11X388.\\$^{*}$Corresponding Author: wzytumbler@scut.edu.cn}
    \thanks{© 2026 IEEE. Personal use of this material is permitted. Permission from IEEE must be obtained for all other uses, in any current or future media, including reprinting/republishing this material for advertising or promotional purposes, creating new collective works, for resale or redistribution to servers or lists, or reuse of any copyrighted component of this work in other works.}
    \end{tabular}
}

\address{
    $^{1}$School of Electrical and Electronic Engineering, Nanyang Technological University, Singapore \\
    $^{2}$Shien-Ming Wu School of Intelligent Engineering, South China University of Technology, China \\
    $^{3}$College of Computer and Information Engineering, Henan Normal University, China \\
    $^{4}$Guangdong Provincial Key Laboratory of Ultra High Definition Immersive Media Technology,\\Peking University Shenzhen Graduate School, China \quad
    $^{5}$VinUniversity, Viet Nam
}

\begin{document}
%
\maketitle
\begin{abstract}

Compared to supervised cross-modal hashing (CMH), unsupervised CMH reduces the reliance on manual labeling by learning binary codes from unlabeled image-text pairs. However, existing unsupervised CMH methods often rely on large-scale image-text pairs, which are costly to collect. To address this limitation, we propose Global-Neighborhood Alignment Hashing (GNAH), a novel approach that preserves the semantic structure of vision–language foundation models within a compact binary Hamming space using only a limited number of image–text pairs.
Specifically, GNAH captures global structural information from the continuous latent space and transfers it into the binary Hamming space through a Prototype-Anchored Global Alignment module. In addition, GNAH extends conventional pairwise contrastive learning by modeling stochastic neighborhood relationships via a Contrastive Stochastic Neighborhood Alignment module, thereby alleviating overfitting to sparse pairwise correlations. Extensive experiments demonstrate that GNAH consistently outperforms existing unsupervised cross-modal retrieval methods under data-constrained settings, offering a practical solution for real-world CMH applications.
\end{abstract}
\section{Introduction}
\label{sec:intro}

The surge in multimedia data, coupled with the advancements in retrieval-augmented generation, has made cross-modal retrieval a compelling topic in both academia and industry. The goal of cross-modal retrieval is to retrieve related instances from one modality ($e.g.,$ image) using a query from another modality ($e.g.,$ text). To achieve efficient cross-modal retrieval, cross-modal hashing (CMH) methods bridge the heterogeneity gap between different modalities by learning a common Hamming space, where instances from different modalities are mapped into compact binary hash codes.

Existing CMH methods can be roughly classified into supervised \cite{jiang2017deep,qin2024deep,huo2023deep,li2023neighborhood} and unsupervised categories \cite{xia2023clip,xi2023unsupervised,zhu2022work,zhuo2022clip4hashing,li2024ckdh,li2023clip,mingyong2023clip}. Unsupervised CMH methods learn hash functions to preserve data correlation without intensive data labeling. Although these methods reduce the dependence on labeled data, they still require a considerable amount of image-text pairs to effectively train the hash functions. These image-text pairs are often scarce and expensive to acquire, particularly when dealing with specialized or private datasets.
Meanwhile, recent vision-language foundation models, such as Contrastive Language-Image Pretraining (CLIP)~\cite{radford2021learning}, have demonstrated remarkable effectiveness in modality alignment, a crucial aspect of cross-modal hashing. This has led to increasing interest in leveraging CLIP for unsupervised CMH \cite{xia2023clip,xi2023unsupervised,zhuo2022clip4hashing,li2024ckdh}. 
However, most existing approaches limit CLIP's role to a feature extractor or focus only on instance-level relationships, failing to exploit its full potential. Given CLIP’s Internet-scale pretraining and proven knowledge transfer capabilities in data-efficient settings for many downstream tasks~\cite{zhou2022learning,li2025class}, a critical question remains: How can we fully harness CLIP for data-efficient unsupervised CMH?

\begin{figure*}[t]
\centering
\includegraphics[width=1.8\columnwidth]{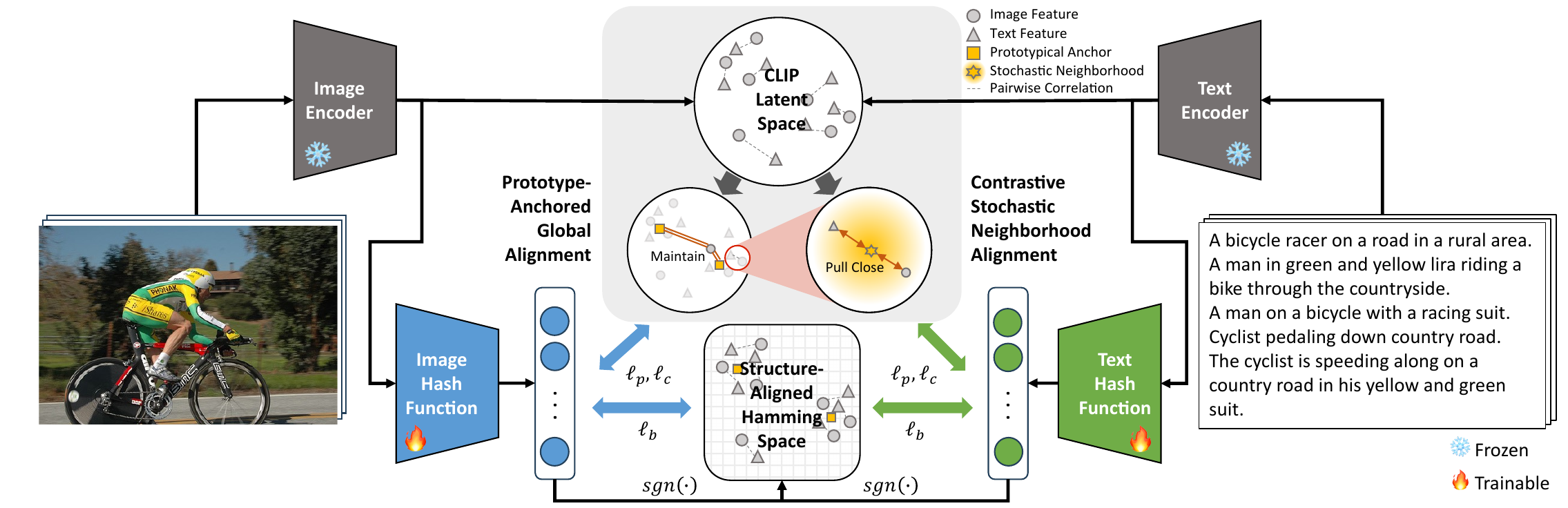} 
\caption{Framework of GNAH. Image and text features from CLIP are mapped into a Hamming space via modality-specific hash functions. Structural information is preserved through Prototype-Anchored Global Alignment for global semantics and Contrastive Stochastic Neighborhood Alignment for local neighborhood priors.}
\label{fig:framework}
\end{figure*}

To bridge this gap, a promising direction is to move beyond simple feature extraction and instance-to-instance correlation learning, instead focusing on leveraging CLIP’s global-neighborhood structural priors to capture semantic relationships across diverse image-text pairs. By transitioning from an instance-to-instance learning paradigm to instance-to-prototype and instance-to-neighborhood approaches, unsupervised CMH can reduce overfitting to individual samples and produce more robust and meaningful hash codes, particularly under data-efficient conditions.
Driven by these motivations, we propose Global-Neighborhood Alignment Hashing (GNAH), a novel approach that enhances data-efficient hashing performance by maintaining the semantic structure of the dataset, facilitating effective knowledge transfer from vision-language foundation models into a compact binary Hamming space.
The main contributions and novelty of this work are summarized as follows:
\begin{itemize}
	\item Our GNAH is a novel method to explore unsupervised cross-modal hashing under data-efficient conditions, reducing the need for well-aligned image-text pairs.
	\item We introduce Prototype-Anchored Global Alignment, which employs representative prototypes as robust semantic references to capture main semantics and maintain global structure in the binary Hamming space.
	\item We develop Contrastive Stochastic Neighborhood Alignment to
    exploit CLIP’s local structural prior by retaining stochastic neighborhood relationships, which mitigates overfitting to limited pairwise correlations.
	\item Extensive experiments show that GNAH outperforms state-of-the-art unsupervised CMH methods and remains competitive with the performance of supervised CMH methods under data-efficient conditions.

\end{itemize}

\section{Methodology}
\subsection{Methodology Overview}

First, inspired by the success of prototype-based methods in data-efficient learning~\cite{snell2017prototypical,wang2019panet,palanisamy2024proto}, we leverage prototypes as robust semantic anchors. Because prototypes offer stable representations that are inherently less vulnerable to noise, they achieve two core goals: capturing the main semantic information for preservation in compact binary codes, and providing resilience to noise through instance-to-prototype relationships. This forms the basis of our Prototype-Anchored Global Alignment (PAGA).

Second, while recent contrastive learning methods~\cite{hu2022UCCH} have demonstrated strong effectiveness in cross-modal retrieval, their strict reliance on pairwise alignment makes them prone to overfitting when pairwise correspondences are scarce. To mitigate this, we excavate CLIP's structural prior and align paired samples with their pairwise neighborhood distributions. Transitioning from instance-to-instance to instance-to-neighborhood alignment forms the foundation of our Contrastive Stochastic Neighborhood Alignment (CSNA), which enhances local neighborhood coherence between the latent and Hamming spaces.

\subsection{Problem Formulation}

Let $D=\{\bold{x}_i, \bold{y}_i\}_{i=1}^n$ denote a cross-modal dataset with $n$ image-text pairs, where $\bold{x}_i \in \mathbb{R}^{d \times 1}$ is the $i^{th}$ instance from image modality, $\bold{y}_i \in \mathbb{R}^{d \times 1}$ is the $i^{th}$ instance from text modality, $d$ is the dimension of image and text features. Simultaneously, we obtain the features of the dataset through the image and text encoders of CLIP as $D_f=\{\bold{f}_i^x, \bold{f}_i^y\}_{i=1}^n$. 
Let $B^x=\{\bold{b}_i^x\}_{i=1}^n$ and $B^y=\{\bold{b}_i^y\}_{i=1}^n$ respectively denote the binary codes of image and text modalities, where $\bold{b}_i^* \in \{-1,+1\}^L$, $* \in \{x,y\}$, and $L$ is the length of hash codes.
We formulate the two hash functions for image and text modalities as $f^x(\bold{f}^x|\Theta ^x)$ and $f^y(\bold{f}^y|\Theta ^y)$ respectively, where $\Theta ^x$ and $\Theta ^y$ are the network parameters to be learned. The outputs of the hash functions are relaxed hash codes defined as $\bold{h}_i^x=f^x(\bold{f}^x_i)$ and $\bold{h}_i^y=f^y(\bold{f}^y_i)$ for the $i^{th}$ image-text pair. The binary hash code of an image-text pair is computed as:
\begin{equation}
    \bold{b}_i^* = \textrm{sgn}(\bold{h}_i^*), \;\; *\in\{x,y\},
\end{equation}
where sgn($\cdot$) is the element-wise sign function.

\subsection{Prototype-Anchored Global Alignment}


\textbf{Alignment with prototypical anchors.}
In this study, we adopt the K-means algorithm to select the prototypical anchors. The K-means cluster centroids are denoted by $\bold{C}=\{\bold{c}_j\}_{j=1}^K$, where the cluster centroids $\bold{C} \in \mathbb{R}^{d \times K}$ are obtained by the K-means algorithm from the multi-modal features $D_f$, and the cluster number $K$ is determined by the Elbow Method~\cite{umargono2020k} to provide a good coverage of the features and represent the global structure.
Meanwhile, in the Hamming space, we initialize $K$ binary prototypical anchors $\bold{C}_b=\{\bold{c}_{bj}\}_{j=1}^K$ corresponding to the prototypical anchors in the latent space, where $\bold{c}_{bj} \in \{-1,1\}^{L \times 1}$. It can be represented in a matrix form as $\bold{C}_b \in \{-1,1\}^{L \times K}$. We enforce $||\bold{h}_i^*||=1$ $(*\in\{x,y\})$ and $||\bold{c}_{bj}||=1$ via $\ell_2$-normalization as in other methods \cite{hu2022UCCH,wu2018unsupervised}. We avoid directly generating binary prototypical anchors by projecting the latent prototypical anchors to the binary Hamming space through hash functions to decouple their relationships, reducing the risk of overfitting to specific feature distributions.
Then, we quantify the relationships between a sample and prototypical anchors as:
\begin{equation}
	\bold{p}^*_i=\text{softmax}\left(\bold{C}^T{\bold{f}^*_i}/\tau\right),
    \quad
    \hat{\bold{p}}^*_i=\text{softmax}\left(\bold{C}_b^T{\bold{h}^*_i}/\tau\right),
\end{equation}
where $\tau$ is the temperature parameter. Since prototypical anchors possess global structural information, we maintain the global structure by aligning the relationships between each sample and prototypical anchors. The loss function for Prototype-Anchored Global Alignment (PAGA) is defined as:
\begin{equation}
	\ell_{p}= \sum\limits_{*} \sum\limits_{i=1}^n \mathcal{D}_{\mathcal{KL}}(\bold{p}^*_i|| \hat{\bold{p}}^*_i) ,
	\label{eq:paga}
\end{equation}
where $\mathcal{D}_{\mathcal{KL}}(\cdot||\cdot)$ denotes the KL divergence.

\textbf{Binary prototypical anchor update.}
As directly optimizing the binary prototypical anchors is an NP-hard problem \cite{shen2015supervised}, we attempt to make the binary prototypical anchors learnable. Specifically, we create prototypical anchors $\{\bold{s}_j\}_{j=1}^K$ using continuous values, and then determine binary prototypical anchors using $\bold{c}_{bj}=\textrm{sgn}(\bold{s}_j)$. This binarization step helps mitigate binarization error, especially given the many samples that gather around these anchors. Inspired by the standard K-means clustering algorithm, the new $j^{th}$ prototypical anchor is calculated as:
\begin{equation}
	\bold{s}_j = \frac{\sum\limits_{i=1}^n (\mathds{1}_{j}(l^x_i)\bold{h}^x_i+\mathds{1}_{j}(l^y_i)\bold{h}^y_i)}{\sum\limits_{i=1}^n (\mathds{1}_{j}(l^x_i)+\mathds{1}_{j}(l^y_i))},
	\label{eq:mix}
\end{equation}
where $l^x_i \in \{1,...,K\}$ denotes the nearest prototypical anchor assignment of the $i^{th}$ instance from image modality, $l^y_i \in \{1,...,K\}$ denotes the nearest prototypical anchor assignment of the $i^{th}$ instance from text modality, and $\mathds{1}_{j}(\cdot)$ is an indicator function defined as:
\begin{equation}
	\begin{array}{l}
		\mathds{1}_{j}(a) = \left\{\begin{matrix} 1 \;\;\;\;\;\;\;\;\;\;\; if \;\;\; a = j
			\\ 0 \;\;\;\;\;\;\;\;\; \textrm{otherwise}.
		\end{matrix}\right.
	\end{array}
\end{equation}

\subsection{Contrastive Stochastic Neighborhood Alignment}


Given an image-text pair $(\mathbf{x}_i, \mathbf{y}_i)$, we define its adaptive neighborhood representation by estimating a neighborhood mean $\boldsymbol{\mu}_i$ and variance $\sigma_i^2$, which serve as statistical descriptors for a Gaussian initialization:

\begin{equation}
\boldsymbol{\mu}_i = \mathrm{norm}(\mathbf{f}^x_i + \mathbf{f}^y_i), \quad \sigma_i^2 = \frac{1}{4} \|\mathbf{f}^x_i - \mathbf{f}^y_i\|_2^2,
\label{eq:mean_variance}
\end{equation}
where $\mathrm{norm}(\cdot)$ denotes the $\ell_2$-normalization.
Using the computed $\boldsymbol{\mu}_i$ and $\sigma_i^2$, we sample from a Gaussian distribution to generate a perturbed representation that captures the neighborhood structure:
\begin{equation}
\tilde{\mathbf{f}}_i = \mathrm{norm}(\boldsymbol{\mu}_i + \sigma_i \boldsymbol{\epsilon}), \quad \boldsymbol{\epsilon} \sim \mathcal{N}(0, \mathbf{I}).
\label{eq:_sample}
\end{equation}

\textbf{Contrastive neighborhood alignment.} 
Once the adaptive neighborhood representations are generated, we project them into the relaxed Hamming space:
\begin{equation}
\tilde{\mathbf{h}}_i^x = f^x(\tilde{\mathbf{f}}_i), \quad \tilde{\mathbf{h}}_i^y = f^y(\tilde{\mathbf{f}}_i).
\label{eq:hamming_projection}
\end{equation}
To enforce alignment within the relaxed Hamming space, we define the contrastive loss function for an image-text pair $(\mathbf{x}_i, \mathbf{y}_i)$ as follows:
\begin{multline}
\ell(\mathbf{x}_i, \mathbf{y}_i) = \\-\log \frac{\exp({\mathbf{h}_i^x}^T \tilde{\mathbf{h}}_i^y / \tau)}{\exp({\mathbf{h}_i^x}^T \tilde{\mathbf{h}}_i^y / \tau) + \sum_{*} \sum_{j=1, j \neq i}^n \exp({\mathbf{h}_i^x}^T \mathbf{h}_j^* / \tau)}.
\label{eq:contrastive_loss}
\end{multline}
This loss function considers the $i$-th instance from the image modality as an anchor and evaluates similarity over both image and text modalities. A symmetric loss term, $\ell(\mathbf{y}_i, \mathbf{x}_i)$, is computed similarly by treating the text modality as the anchor.
The loss function for Contrastive Stochastic Neighborhood Alignment (CSNA) is defined as:
\begin{equation}
\ell_c = \sum_{i=1}^{n} \ell(\mathbf{x}_i, \mathbf{y}_i) + \ell(\mathbf{y}_i, \mathbf{x}_i).
\label{eq:final_contrastive}
\end{equation}

\begin{figure*}[htbp]
    \centering

    \includegraphics[width=0.24\textwidth]{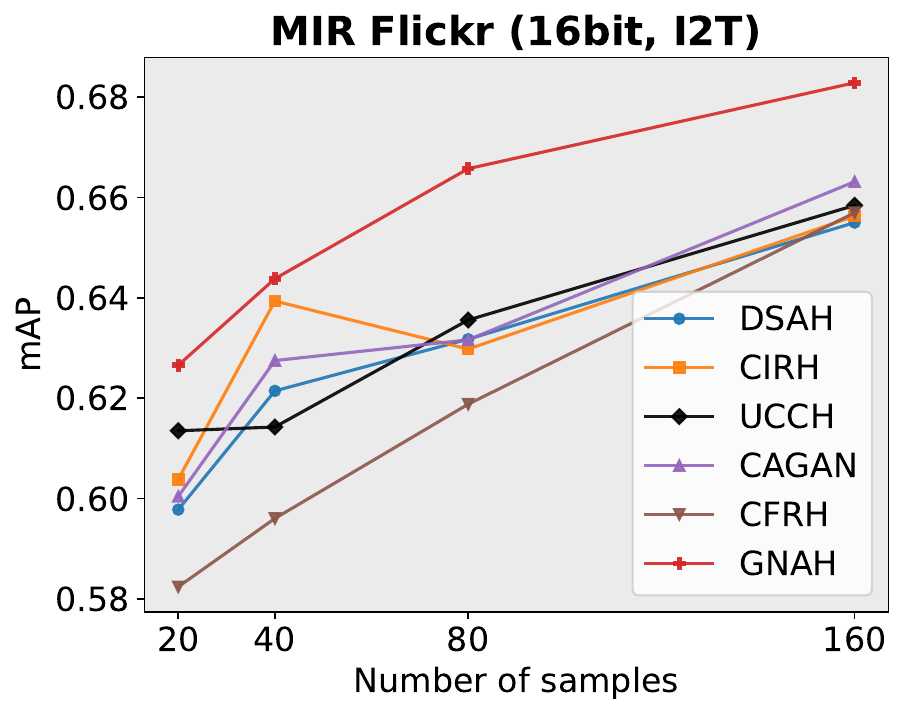}
    \includegraphics[width=0.24\textwidth]{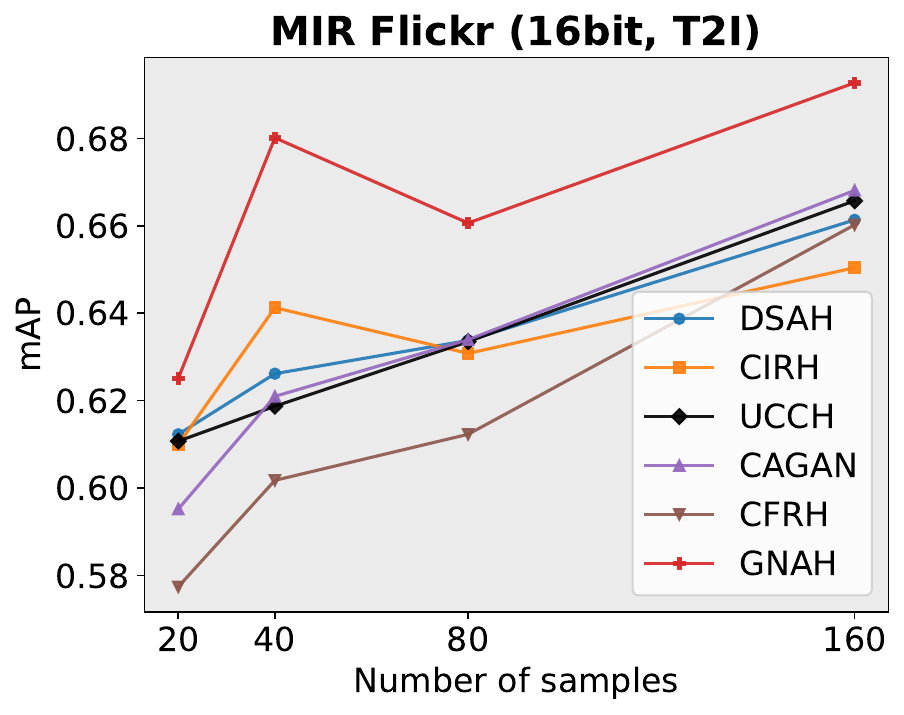}
    \includegraphics[width=0.24\textwidth]{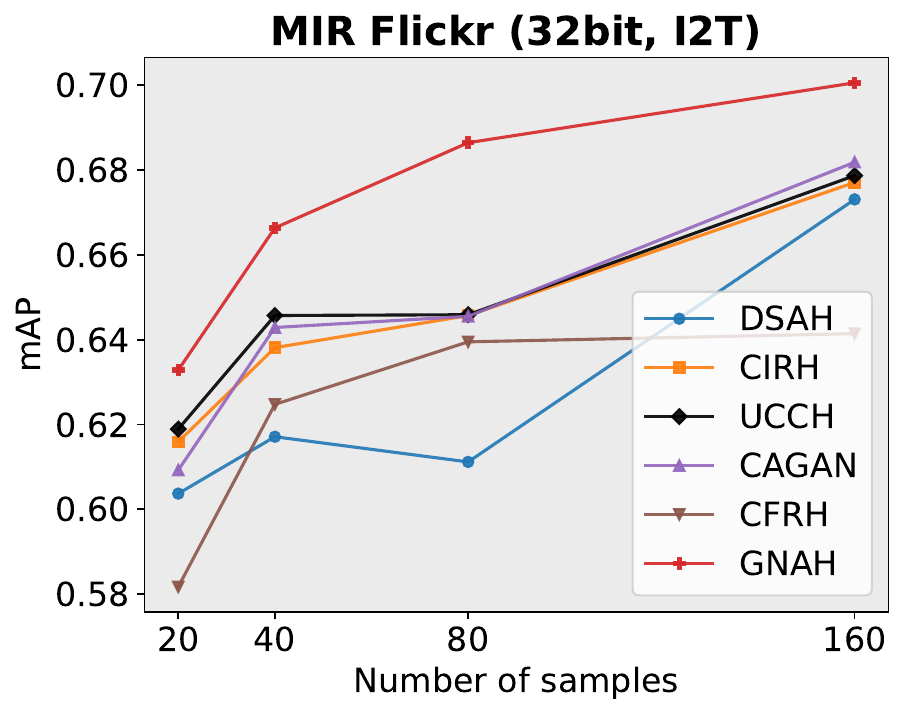}
    \includegraphics[width=0.24\textwidth]{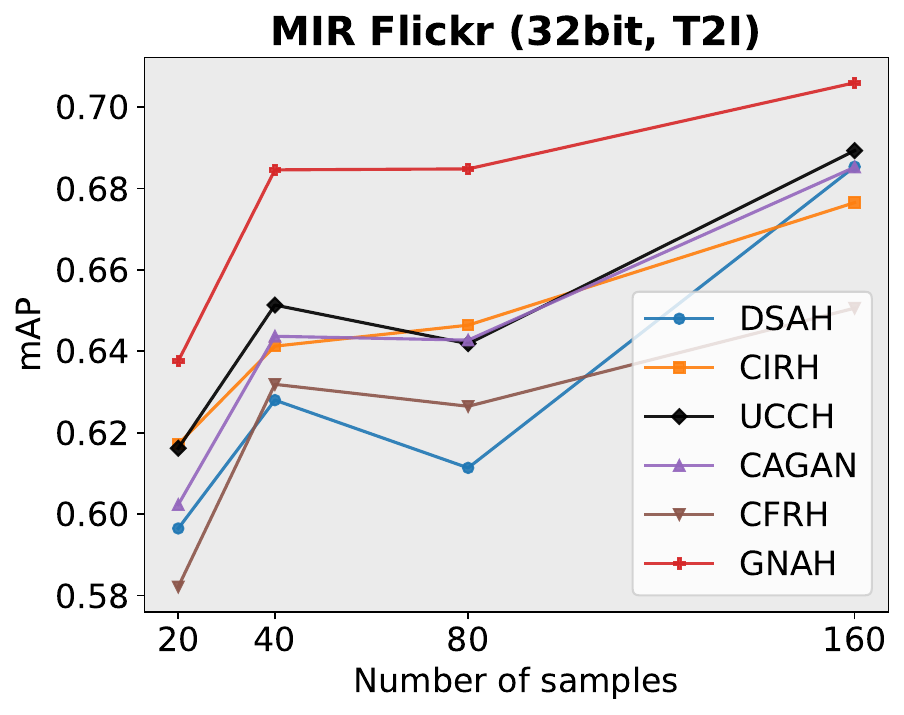}
    \includegraphics[width=0.24\textwidth]{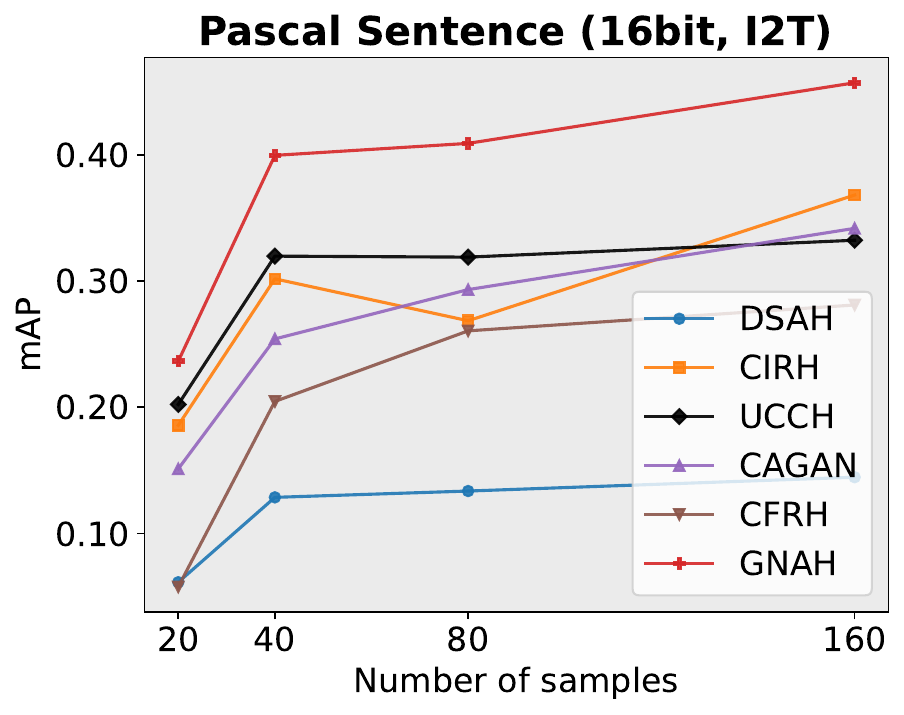}
    \includegraphics[width=0.24\textwidth]{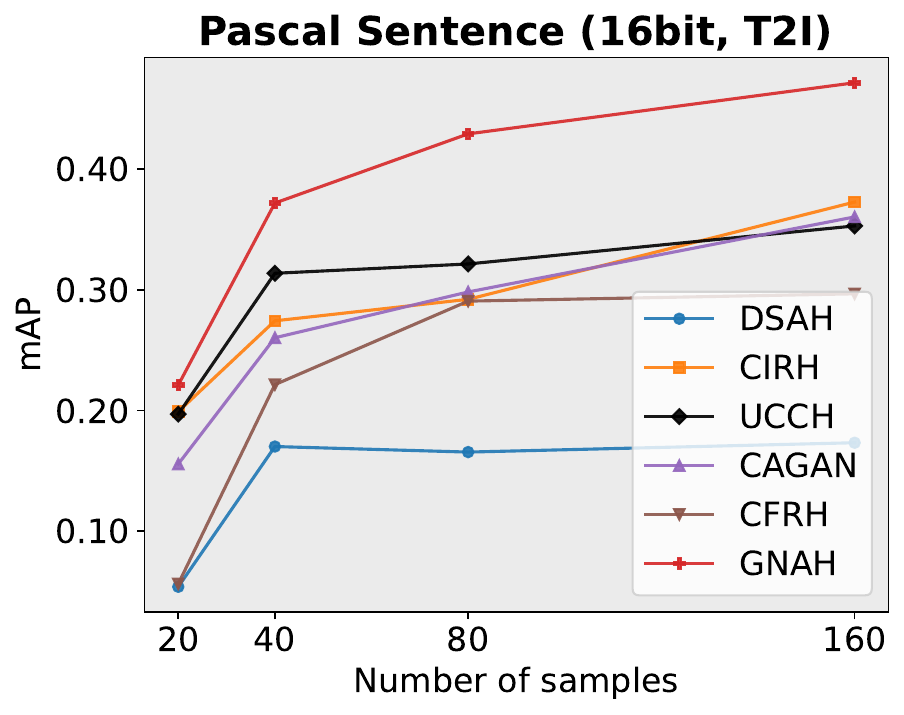}
    \includegraphics[width=0.24\textwidth]{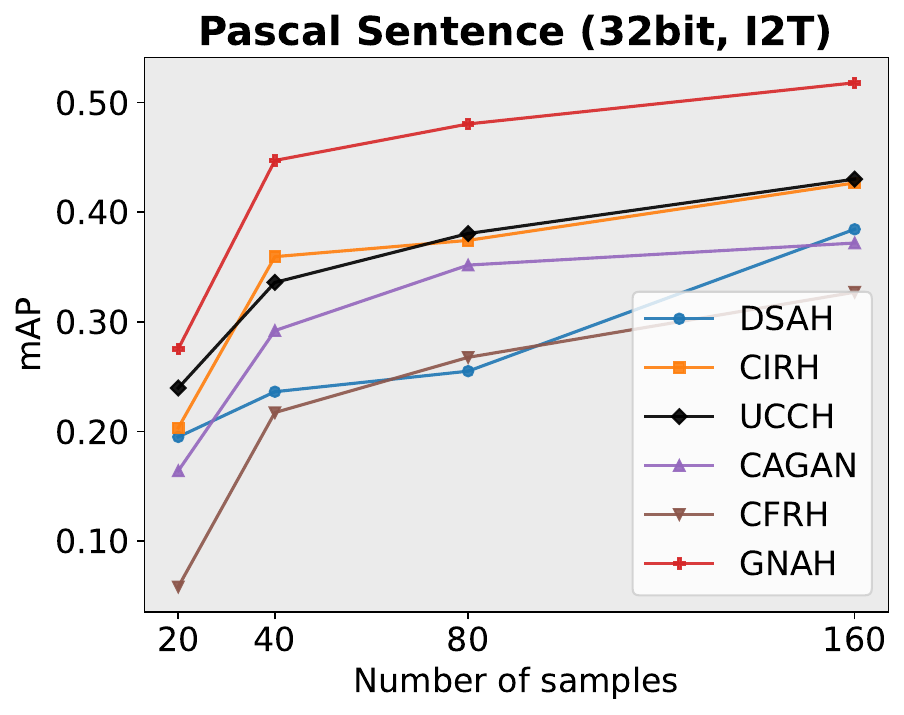}
    \includegraphics[width=0.24\textwidth]{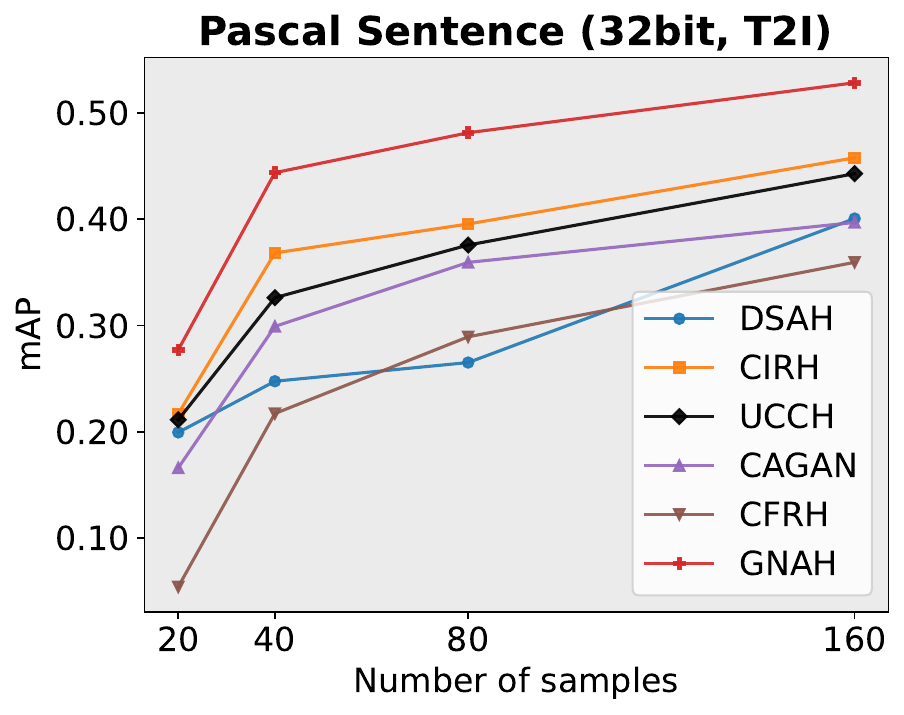}
    \includegraphics[width=0.24\textwidth]{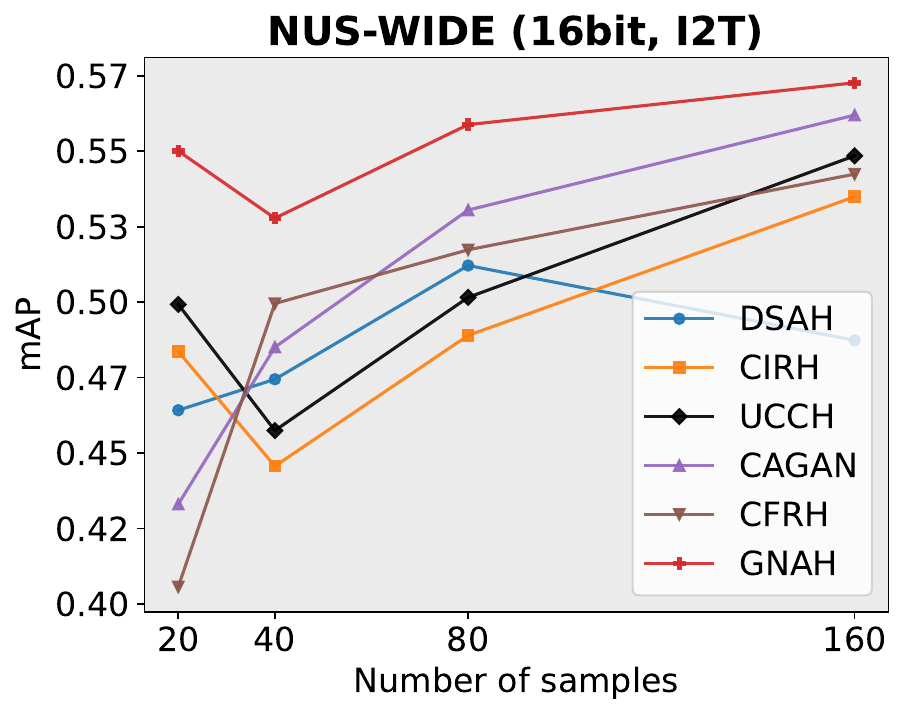}
    \includegraphics[width=0.24\textwidth]{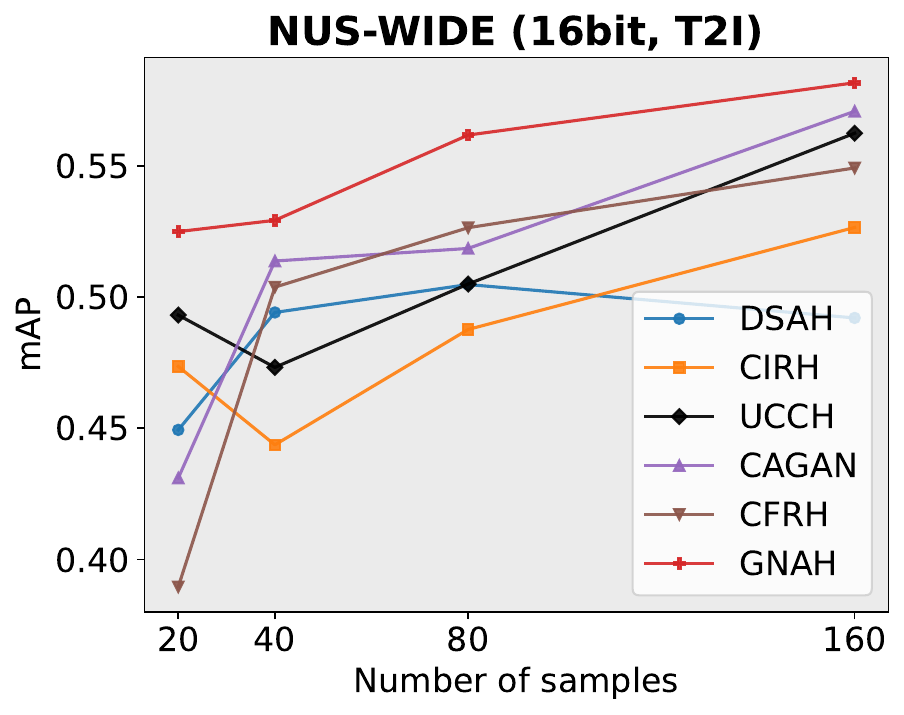}
    \includegraphics[width=0.24\textwidth]{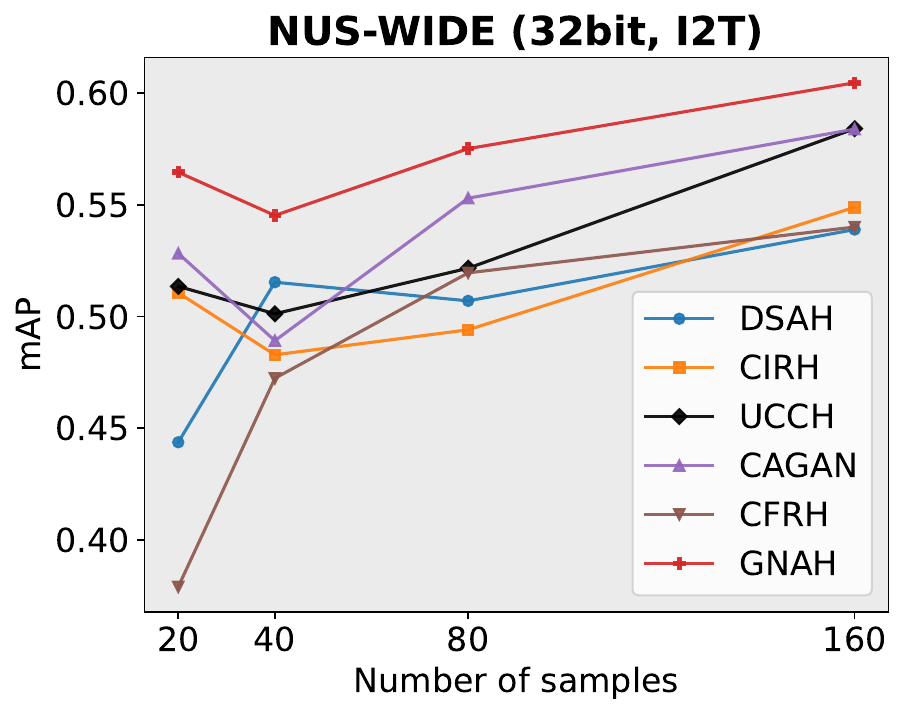}
    \includegraphics[width=0.24\textwidth]{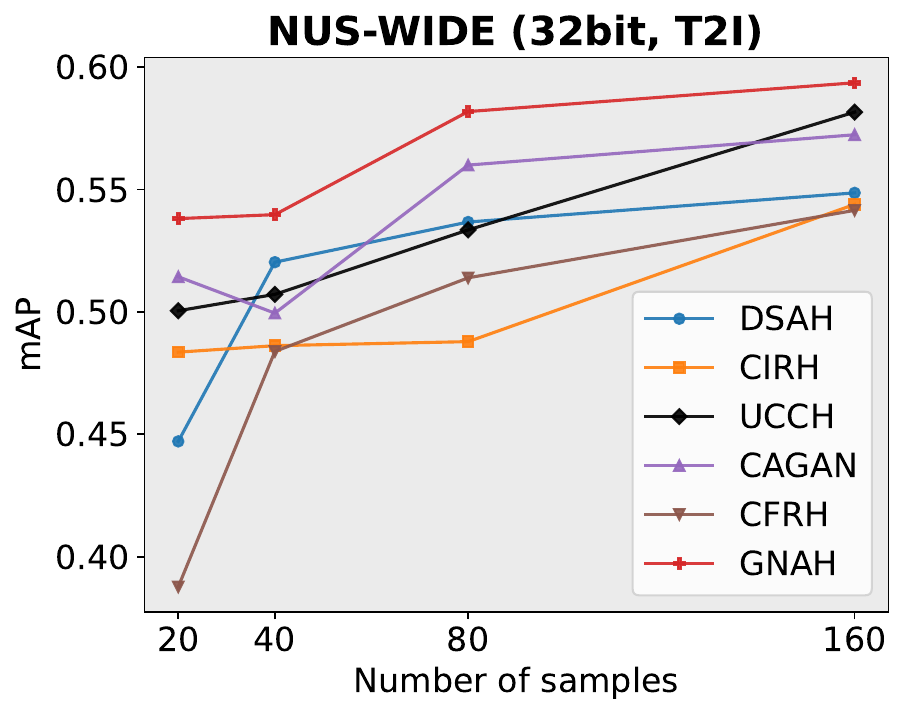}

    \caption{Retrieval performance of unsupervised CMH methods through data-efficient learning at 16 and 32 bits. ``I2T" denotes image-to-text retrieval and ``T2I" denotes text-to-image retrieval.}
    \label{fig:unsuResults}
\end{figure*}

\subsection{Binarization and Overall Optimization}
\textbf{Binarization error reduction.}
Similar to the prior method~\cite{jiang2017deep}, to mitigate binarization errors caused by continuous relaxation and enhance the representational integrity of our hash functions, we incorporate a binarization loss function defined as:
\begin{equation}
    \ell_{b}= \sum\limits_{*} \sum\limits_{i=1}^n \|\mathbf{h}_i^*-\mathbf{b}_i^*\|^2_2.
	\label{eq:quant}
\end{equation}

\textbf{Overall objective.}
Finally, by combining  $\ell_{p}$, $\ell_{c}$, and $\ell_{b}$, we formulate the overall objective of GNAH as:
\begin{equation}
	\ell = \beta \exp(-\gamma t) \ell_p+(1-\beta \exp(-\gamma t))\ell_c + \ell_{b},
	\label{eq:obj}
\end{equation}
where $0 \le \beta \le 1$ is a hyper-parameter to control the strength of the global alignment, $\gamma \geq 0$ determines the exponential decay of PAGA, and $t$ represents the number of epochs. 
We utilize a curriculum learning approach, where PAGA initially establishes a global structural foundation, then gradually decays to prioritize local refinement via CSNA.


\section{Experiments}

\subsection{Implementation Details}

Our method is evaluated using three widely used datasets: MIR Flickr~\cite{huiskes2008mir}, Pascal Sentence~\cite{rashtchian2010collecting}, and NUS-WIDE~\cite{chua2009nus}.
For all datasets, training samples are randomly drawn from the retrieval sets with a fixed random seed 42. To assess the model’s performance under varying levels of data scarcity, we conduct evaluations using 20, 40, 80, and 160 training sample pairs per dataset.
Following the convention \cite{hu2022UCCH}, mean average precision (mAP) is adopted for performance evaluation.

We compare our method against five unsupervised CMH methods, CIRH~\cite{zhu2022work}, DSAH~\cite{yang2020deep}, UCCH~\cite{hu2022UCCH}, CAGAN~\cite{li2023clip}, and CFRH~\cite{mingyong2023clip}. We also compare our method against two recent supervised CMH methods, DNPH~\cite{qin2024deep} and DSPH~\cite{huo2023deep}. To ensure a fair comparison, all methods use CLIP's image and text encoders as feature extractors. 
The hyperparameters of GNAH and other methods are tuned based on the dataset or set as default values. The GNAH model is trained using the Adam~\cite{kingma2014adam} optimizer for 500 epochs with a batch size of 100. The learning rate adopted for training is 0.0001.
Comparisons are conducted at 16 and 32 bits.

\begin{table*}[h]
    \footnotesize
    \centering
    \caption{Performance of GNAH and its variants at 32 bits with 80 training sample pairs.}
    \begin{tabular}{l|cccccccccccccc}
        \toprule
        Dataset & \multicolumn{2}{c}{Full GNAH} & \multicolumn{2}{c}{w/o $\ell_p$} & \multicolumn{2}{c}{w/o $\ell_c$} & \multicolumn{2}{c}{w/o $\ell_{b}$} & \multicolumn{2}{c}{GNAH-A} & \multicolumn{2}{c}{GNAH-B} & \multicolumn{2}{c}{GNAH-C} \\
        \cmidrule(lr){2-3} \cmidrule(lr){4-5} \cmidrule(lr){6-7} \cmidrule(lr){8-9} \cmidrule(lr){10-11} \cmidrule(lr){12-13} \cmidrule(lr){14-15}
        & I2T & T2I & I2T & T2I & I2T & T2I & I2T & T2I & I2T & T2I & I2T & T2I & I2T & T2I \\
        \midrule
        MIR Flickr     & \textbf{0.686} & \textbf{0.685} & 0.669 & 0.669 & 0.576 & 0.580 & 0.682 & 0.681 & 0.656 & 0.647 & 0.681 & 0.678 & 0.671 & 0.677 \\
        Pascal Sentence & \textbf{0.480} & \textbf{0.481} & 0.448 & 0.449 & 0.123 & 0.119 & 0.468 & 0.469 & 0.337 & 0.369 & 0.447 & 0.473 & 0.453 & 0.454 \\
        NUS-WIDE      & 0.575 & 0.582 & 0.571 & 0.571 & 0.500 & 0.475 & 0.568 & 0.573 & 0.564 & 0.554 & \textbf{0.580} & 0.574 & 0.578 & \textbf{0.587} \\
        \bottomrule
    \end{tabular}

    \label{tab:variants}
\end{table*}

\subsection{Data-Efficient Learning for Cross-Modal Hashing }

\textbf{Comparison with unsupervised CMH methods.}
 As shown in Fig.~\ref{fig:unsuResults}, GNAH consistently outperforms all baselines across datasets. For instance, at 32 bits with 80 training samples, GNAH achieves 0.686 (I2T) and 0.685 (T2I) on MIR Flickr, exceeding the next-best methods by 4.1\% and 3.8\%. On Pascal Sentence, GNAH reaches 0.480 (I2T) and 0.481 (T2I), outperforming the next-best methods by 10.0\% and 8.6\%. On NUS-WIDE, GNAH achieves 0.575 (I2T) and 0.582 (T2I), surpassing CAGAN by 2.2\% for both tasks.

\begin{figure}[]
    \centering

    \includegraphics[width=0.235\textwidth]{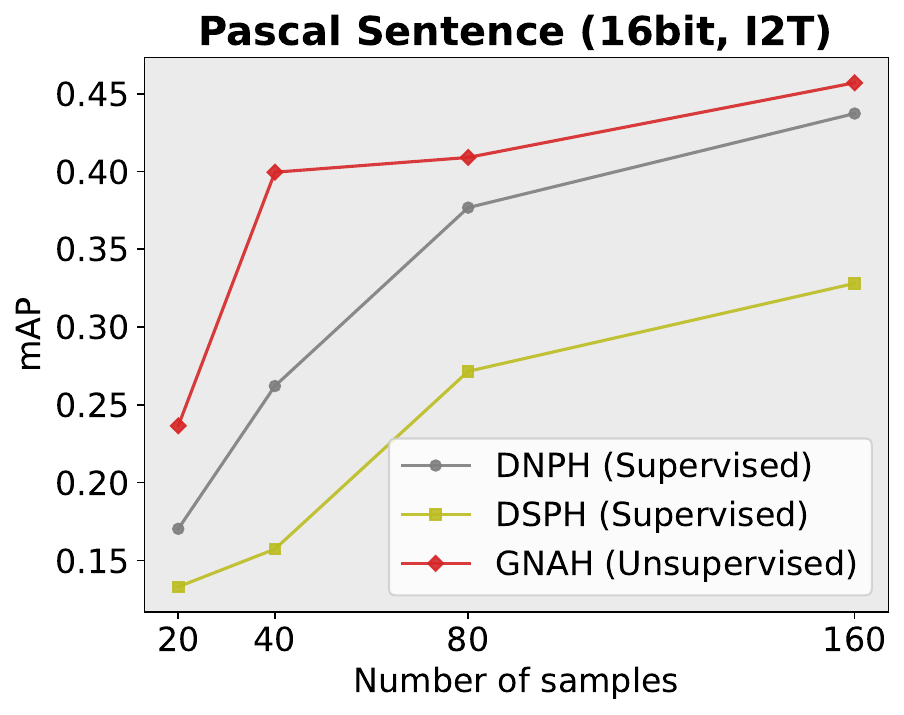}
    \includegraphics[width=0.235\textwidth]{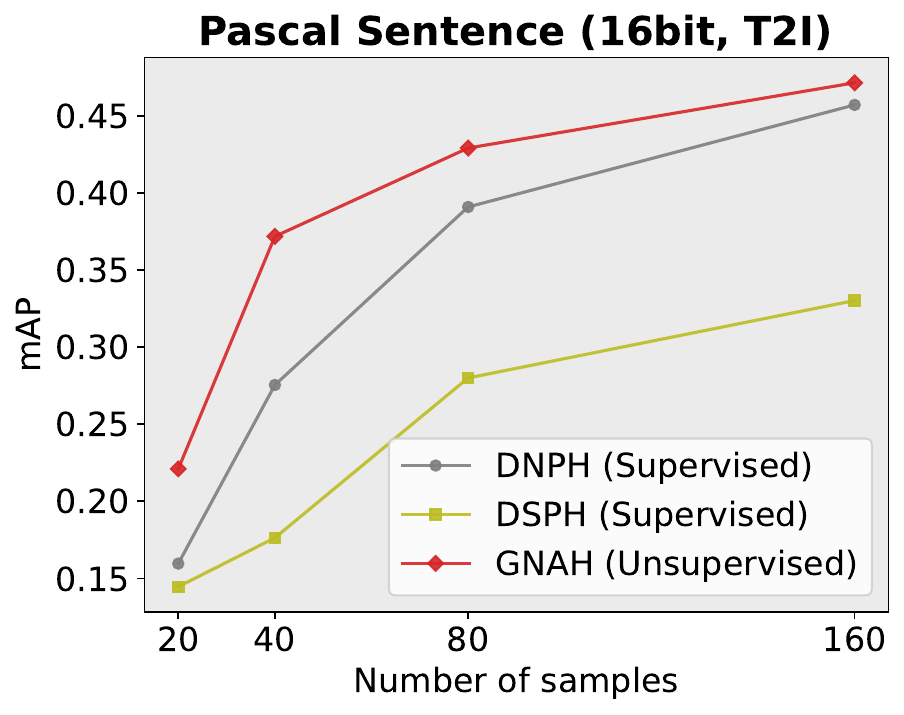}
    \includegraphics[width=0.235\textwidth]{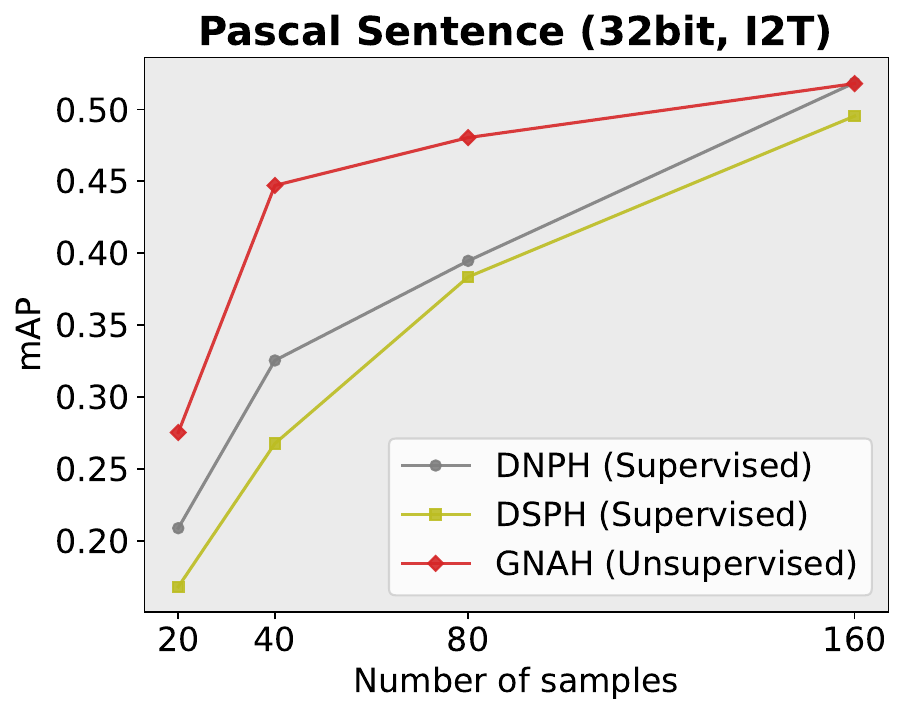}
    \includegraphics[width=0.235\textwidth]{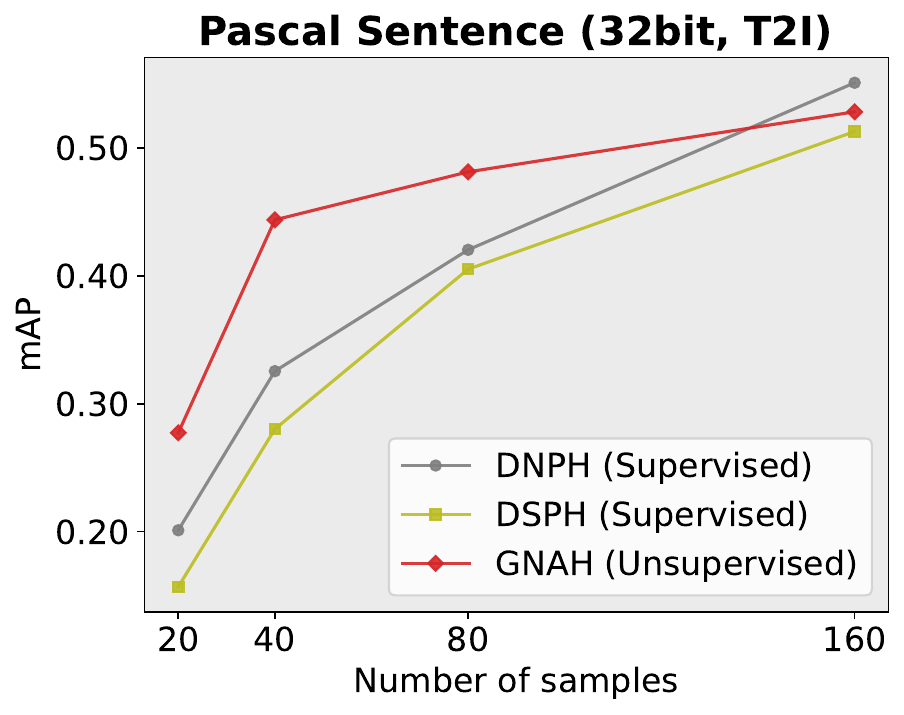}

    \caption{Retrieval performance of GNAH and supervised CMH methods through data-efficient learning at 16 and 32 bits. }
    \label{fig:suResults}
\end{figure}

\textbf{Comparison with supervised CMH methods.}
We further compare GNAH with two supervised CMH methods on Pascal Sentence. As shown in Fig.~\ref{fig:suResults}, GNAH consistently outperforms supervised baselines with 20–80 sample pairs despite using no labels. Although supervised methods improve with 160 samples as more labeled data reduces overfitting, our unsupervised approach remains competitive without requiring any manual annotations.

\subsection{Ablation Study}
\begin{figure}[]
    \centering
    \hfill
    \includegraphics[width=0.48\textwidth]{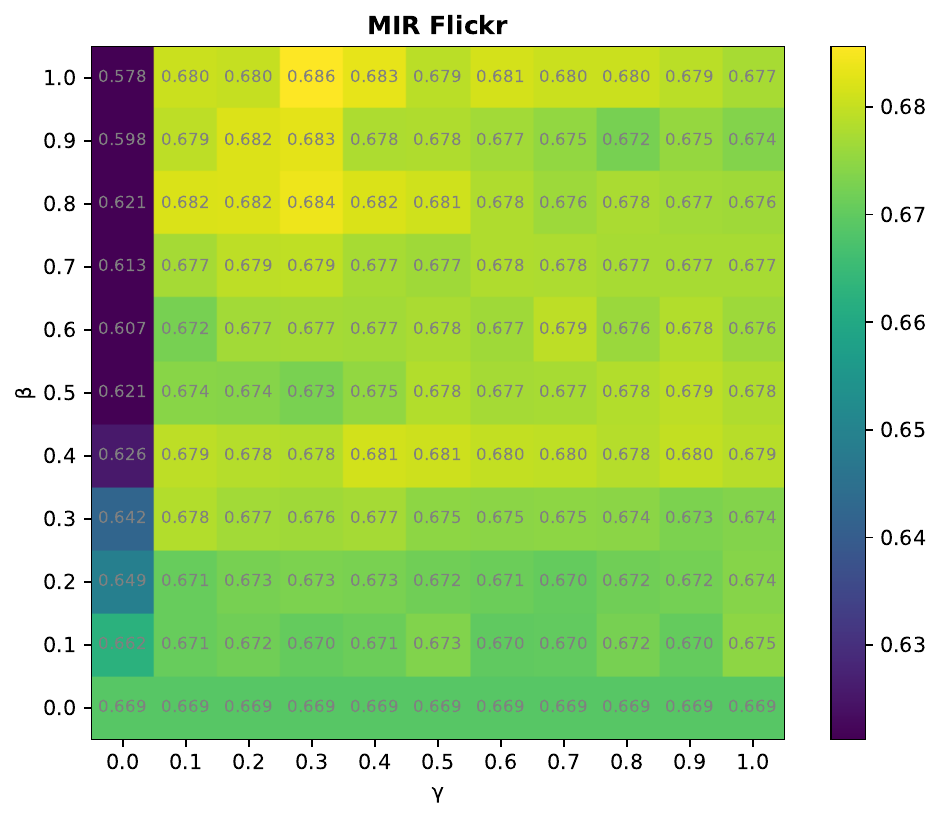}\hfill
    \hfill
    \caption{Effect of $\beta$ and $\gamma$ at 32 bits with 80 training sample pairs. Results are averaged between I2T and T2I.}
    \label{fig:effect}
\end{figure}

\textbf{Effect of $\beta$ and $\gamma$.}
To analyze how the global alignment weight $\beta$ and exponential decay factor $\gamma$ affect performance, we conduct a grid search on MIR Flickr. As shown in Fig.~\ref{fig:effect}, the average I2T and T2I mAP at 32 bits (with 80 training pairs) reveals clear interactions between these hyperparameters.
When $\beta = 0$, the model relies solely on CSNA, yielding lower performance and indicating that CSNA alone cannot sufficiently capture dataset-level semantics. When $\gamma = 0$, PAGA does not decay, which also harms performance, since PAGA is not a perfect modality aligner and should not dominate throughout training. Allowing PAGA to shape the initial global structure, then shifting focus to CSNA for local refinement, yields better results.

\textbf{Variants of GNAH.} To investigate the impact of different components and explore alternative configurations, we construct several variants of GNAH, including: (1) GNAH without $\ell_p$, (2) GNAH without $\ell_c$, (3) GNAH without $\ell_{b}$, (4) GNAH-A: replacing the proposed contrastive loss with a traditional pairwise contrastive loss for modality alignment, (5) GNAH-B: relaxing the binary prototypical anchors by removing the binarization step, and (6) GNAH-C: generating binary prototypical anchors by directly projecting the latent prototypical anchors through hash functions and subsequently binarizing the relaxed binary prototypical anchors.
Table~\ref{tab:variants} reports the performance of GNAH and its variants at 32 bits with 80 training pairs. The full model consistently outperforms variants (1)–(3), highlighting the importance of each component. Notably, removing $\ell_c$ results in a sharp performance drop, particularly on Pascal Sentence (declining from 0.480 to 0.123 in I2T and from 0.481 to 0.119 in T2I). This degradation happens because PAGA focuses mainly on capturing global semantics while overlooking modality alignment and local neighborhood structure. 


\section{Conclusion}

In this paper, we propose Global-Neighborhood Alignment Hashing (GNAH) for unsupervised data-efficient cross-modal retrieval. By integrating Prototype-Anchored Global Alignment and Contrastive Stochastic Neighborhood Alignment, GNAH effectively transfers semantic structures from foundation models into a compact Hamming space while mitigating overfitting in low-data regimes. Experimental results across multiple benchmarks demonstrate that GNAH outperforms state-of-the-art unsupervised methods and remains competitive with supervised approaches. 

\ninept            
\bibliographystyle{IEEEbib}
\bibliography{strings,refs}

\end{document}